\documentclass[12pt,preprint]{aastex}

\newcommand{\twcoto}{$^{12}$CO $J$=2$-$1}
\newcommand{\twcooz}{$^{12}$CO $J$=1$-$0}

\newcommand{\joz}{$J$=1$-$0}
\newcommand{\etal}{et al.}
\newcommand{\eg}{e.g.}
\newcommand{\ie}{i.e.}
\newcommand{\kms}{km s$^{-1}$}

\newcommand{\msun}{M$_\odot$}
\newcommand{\h}{$^{\rm h}$}
\newcommand{\m}{$^{\rm m}$}

\shorttitle{CO in Double Barred Galaxies}
\shortauthors{Petitpas \& Wilson}

\begin{document}

\title{Molecular Gas in Candidate Double Barred Galaxies I. The
Diverse Morphology and Dynamics of NGC 2273 and NGC 5728}
\author{Glen R. Petitpas\altaffilmark{1} \& Christine D. Wilson}

\affil{Department of Physics and Astronomy, McMaster University, 1280
Main Street West, Hamilton Ontario,
Canada L8S 4M1}
\email{petitpa@physics.mcmaster.ca \\ wilson@physics.mcmaster.ca}

\altaffiltext{1}{Current address: Department of Astronomy, University of Maryland, College Park MD, USA, 20742}

\shorttitle{CO Maps of Candidate Double Barred Galaxies}
\shortauthors{Petitpas \& Wilson}

\begin{abstract}

Double bars have been proposed as a means of transporting molecular gas
past inner Lindblad resonances into the nuclear regions, where it can
fuel active or starburst nuclei. Thus far, the existence of double bars
has been determined predominantly through analysis of near
infrared images, which can tell us little about the dynamics and inflow
rates of these systems. We have observed two double bar galaxy
candidates (NGC 2273 and NGC 5728) in \twcooz\ with the Owens Valley
Radio Observatory Millimeter Array. Despite the similarity in the near
infrared images of these galaxies, we see rather different nuclear
morphologies in the CO maps. NGC 2273 shows evidence of a nuclear gas
bar, aligned with the nuclear stellar bar seen in the near infrared
images. Both the nuclear gaseous and stellar bars are misaligned from
the large scale bar by $\sim$90\arcdeg, which also allows the
possibility that both are the result of stars and gas populating the
$x_2$ orbits of the primary bar. Estimates using dynamical friction
arguments and star formation rates suggest significant gas inflow
rates along the nuclear bar of NGC 2273. Conversely, NGC 5728 does not
show any evidence for a nuclear molecular bar, but shows an arc of CO
clumps that peaks just to the south-west of the dynamical center and
curves to the south-east where it follows the dust lane to the south.
Models of double-barred galaxies suggest that these galaxies should
contain large amounts of molecular gas in their nuclei. Our
calculations suggest that both galaxies contain sufficient amounts of
gas in their nuclei, but only NGC 2273 shows evidence for a nuclear gas
bar. This may be the result of past episodes of star formation
exhausting and dispersing the nuclear gas of NGC 5728, but is more
likely evidence that NGC 5728 has undergone a minor merger event.

\end{abstract}


\section{Introduction}

The nuclei of barred spiral galaxies are often the setting for
extraordinary events such as starbursts, molecular rings, inflow, and
even Seyfert activity. The need to understand the mechanisms driving
these phenomena has inspired a great number of observations and
computer simulations. Models suggest that bars in galaxies can drive
molecular gas into the nuclear regions where it can fuel the
vigorous star formation activity that would otherwise exhaust the
molecular gas content on timescales much shorter than observed
\citep[\eg,][]{com94}. Bars can only drive molecular gas inward until
it approaches the Inner Lindblad Resonance (ILR; see
Fig.~\ref{schematic}), where it will halt its inflow and
accumulate into a ring. To overcome this, \citet{shl89} proposed that
the ring may become unstable and form a secondary bar inside the radius
of the ILR, which could allow gas to reach much farther into the
nucleus and possibly be the driving mechanism behind Seyfert nuclei.

Recent near infrared (NIR) surveys reveal isophote twists in the
central regions of barred galaxies which may be the signature of this
`bar within a bar' \citep[\eg,][]{mul97}. There are three mechanisms
which can account for NIR isophotal twists \citep{elm96}. The first
mechanism (hereafter called Model 1), proposed by \citet{sha93},
suggests the isophote twists are the result of twisting of the primary
bar triggered by a dissipative gaseous component and misaligned from
the primary bar. Their numerical simulations suggest that in the
presence of two ILRs a nuclear ring can become elongated perpendicular
to the primary bar (along the $x_2$ orbits). Gas dissipation causes the
inner part of the perpendicular gaseous ring to become more aligned
with the primary bar, resulting in the appearance of an elongated
nuclear ring that leads the primary bar. This gas ring exerts a torque
on the stellar component of the bar, pulling it out of alignment also.
The whole system would then rotate with the same angular frequency,
with the inner gaseous ring and inner part of the primary bar leading
the outer parts of the primary bar. The second mechanism (hereafter
Model 2) suggests that the twists are the result of a kinematically
distinct secondary bar \citep{fri93}. Their N-body simulations (with
stars and gas) suggests gas inflow along the primary bar can accumulate
enough mass that the inner part of the gas bar can become nearly
self-gravitating and decouple from the primary bar. The secondary bar
may rotate with a pattern speed of up to 6 times that of the primary
bar. The third mechanism (Model 3) suggests that the NIR isophote
twists may be the result of a triaxial stellar bulge \citep{kor79}.

Detailed analyses of NIR images reveal that, in many cases, these
isophote twists must be caused by the existence of true nuclear stellar
bars \citep{fri96,jun97}. Typically, nuclear stellar bars are
identified by two distinct peaks in the plots of ellipticity versus
radius, as well as a sudden change in the position angle of the
isophotes at the end of the nuclear bar \citep[unless the nuclear and
large scale bars are aligned;][]{woz95}. Unfortunately, even at NIR
wavelengths, dust obscuration and contamination by young red
supergiants in regions of star formation in the nuclei of these
galaxies with isophote twists can make it difficult to identify
nuclear bars unambiguously \citep[\eg,][]{fri96,rho98}. Fortunately,
since molecular gas observations in the millimeter and submillimeter
regime are not prone to dust extinction, observations of \twcooz\ can
be used to probe the morphology and dynamics of the molecular gas in
these nuclei.

The three competing explanations for the NIR isophote twists can also
be tested using these molecular gas observations. Model 1 would exhibit
an secondary gaseous bar that may lead the secondary stellar bar
slightly, but has the same rotation speed as the primary bar. Model 2
would show a gaseous secondary bar that is rotating with a different
pattern speed than the primary bar. Thus, molecular gas dynamics should
allow us to distinguish between these models. Model 3 is associated
with the stellar bulge. Since there is very little gas in the bulge
compared to the disk of a galaxy, the CO maps need not resemble the
NIR images. If the CO maps do share similar features with the NIR images,
it may suggest that the isophote twists are related to the disk
of the galaxy, and not the bulge. Both the first and second models were
only able to re-create relatively long lived (\ie\ a few $\times 10^8$
yrs) nuclear NIR isophote twists in models that contained sufficient
amounts of molecular gas (initially 2-10\% gas globally; higher in the
nuclear regions at later stages of evolution). Without these high gas
fractions, there is not enough dissipation in the models, and the
observed structures do not last long enough to be as common as they are
observed to be \citep{sha93,fri93}.

Previous CO studies of galaxies with nuclear stellar bars have revealed
a variety of CO morphologies, but most show some evidence for a nuclear
gas bar. Perhaps the current best example of a galaxy with nuclear
gaseous and stellar bars is NGC 2782 \citep{jog99}. The NIR images of
this galaxy show evidence for a nuclear stellar bar in the ellipticity
of the isophotes. The CO map for this galaxy resembles a twin peaked
morphology, with the twin peaks of CO emission aligned with the nuclear
stellar bar. The CO data show kinematic evidence for inflow along the
nuclear bar. Estimates of the inflow rates for NGC 2782 suggest values
that are high enough to sustain the star formation rates of $\sim$ 3 -
6 \msun\ yr$^{-1}$ for approximately $5 \times 10^8$ years.

Previous studies of M100 have shown the existence of both a nuclear
stellar bar and a nuclear gas bar. \citet{gar98}
modeled the large and small scale HI and CO observations using a
variety of numerical models and found that the best fit to the
observations was for the case of a kinematically distinct nuclear bar
rotating at nearly seven times the pattern speed of the large scale
bar. The kinematically distinct nuclear bar was needed in the models to
create enough inflow to reproduce the high concentration of gas seen in
the nuclear disk of M100. Unfortunately, no estimates of the inflow
rate are given.

Studies of NGC 4736 by \citet{won00} show a nuclear gas bar roughly
aligned with the nuclear stellar bar seen in the NIR images by
\citet{sha93}. The CO maps show evidence for inflow rates of $\sim$ 2
\msun\ yr$^{-1}$ for the outer bar, but no estimates for inflow rates
along the nuclear bar are given. Presumably, these rates would have to
be on the order of at least 0.2 \msun\ yr$^{-1}$ to sustain the nuclear
star formation rates.

In other cases there is evidence for a nuclear gas bar where there is
no evidence for a nuclear stellar bar. For example, NGC 3351
\citep{dev92} was the first galaxy to show a molecular gas bar
perpendicular to the large scale stellar bar. Unfortunately,
\citet{dev92} did not resolve the bar across its minor axis, so they
are unable to determine if there are kinematic signatures of inflow.
Later studies of this galaxy suggest that the nuclear gas bar is 
simply the gas collecting at ILRs \citep{ken92}. 

This paper presents observations taken with the Owens Valley Radio
Observatory Millimeter Array of two galaxies that exhibit NIR isophote
twists: NGC 2273 and
NGC 5728. NGC 2273 is a SB(r)a galaxy that has a Seyfert 2 nucleus,
nuclear star formation \citep{mul96}, and a nuclear ring of dust
\citep[$r \approx 5''$;][]{yan99}. It has a recession velocity of 1841
\kms\ \citep{dev91} which implies that it is 24.5 Mpc away (assuming
H$_0$ = 75 \kms\ Mpc$^{-1}$). It also has three outer rings which
appear to be made by separate sets of spiral arms at $r \approx
0\farcm4$, $1\farcm1$ and $1\farcm0$. HI observations indicate that it
is not very gas rich for a barred spiral galaxy with such a wide
variety of nuclear activity \citep{van91}. This observation is
supported by the \twcooz\ spectra taken by \citet{you91}, who find very
narrow CO linewidths and conclude that most of the molecular gas must
be contained in the central few arcseconds of the nuclei of NGC 2273.
Color maps of the inner regions of NGC 2273 suggest that there is a
reddened ring of dusty material ($r \sim$ 5\arcsec) surrounding a
region of high ionization \citep{yan99}. NGC 5728 is a southern
hemisphere barred spiral galaxy classified as SAB(r)a that also
contains a Seyfert 2 nucleus. Its recession velocity of 2788
\kms\ \citep{dev91} suggests it is located at a distance of 37.2 Mpc.
Color maps indicate the galaxy contains two blue rings of recent star
formation, one in the nucleus \citep[$r \approx 5''$;][]{wil93} and one
near the primary bar ends \citep[$r \approx 55''$;][]{sch88}. Two dust
lanes emerge from just outside the nuclear ring and run parallel to the
primary bar of the galaxy. Table \ref{table} summarizes the properties
and adopted parameters for these galaxies. These two galaxies were
chosen as targets for this study because of the similarity in their
nuclear and large scale morphologies; both galaxies are classified as
Seyfert 2 galaxies and contain nuclear rings (which suggest the
presence of at least one ILR), and, most importantly, both galaxies
exhibit the NIR isophote twists which may be the signature of a nuclear
bar as discussed above.

In \S\ref{obs} we discuss the observations and data reduction
techniques. In \S\ref{morphanddyn} we discuss the molecular gas
distribution and dynamics. In \S\ref{mass} we discuss the molecular gas
mass determined from the \twcooz\ flux in comparison with the amount of
molecular gas required by the models to produce the observed features.
In \S\ref{inflow} we search for signs of inflow in NGC 2273 and compare
these with star formation rates. In \S\ref{prev} we compare our results
to models of double barred galaxies and previous observations of these
galaxies. In \S\ref{other} we discuss our observations in relation to
observations of other galaxies thought to contain nuclear stellar and
gaseous bars. The paper is summarized in \S\ref{conc}.

\section{Observations and Data Reduction \label{obs}}

We have observed the barred galaxies NGC 2273 and NGC 5728 in \twcooz\
(115.3 GHz) using the Owens Valley Radio Observatory Millimeter Array.
For NGC 2273 we have 3 tracks, two in the low-resolution (L)
configuration, and one in the high-resolution configuration (H). For
NGC 5728 we have four tracks (two in H, one in L, and one in the
equatorial (E) configuration).  Preliminary calibration was done using
the calibration package MMA \citep{sco93}. Only data with a coherence
$>$0.5 were used. The quasars 0642+449 and 1334$-$127 were used for
gain calibration and Neptune and 3C273 were used for flux calibration.

For NGC 2273, we used $\alpha$(2000) = 06\h50\m08\fs7, $\delta$(1950) =
+60\arcdeg50\arcmin45\farcs1 for the pointing center (note: this is
slightly different from the galactic center) and the spectrometer was
centered at $V_{\rm lsr}$ = 1841 \kms. For NGC 5728 we used the
pointing center of $\alpha$(2000) = 14\h42\m24\fs0, $\delta$(2000) =
$-$17\arcdeg15\arcmin10\farcs8 and a recession velocity of $V_{\rm
lsr}$ = 2930 \kms. In both cases, the spectrometer had a bandwidth of
240 MHz which corresponds to a velocity coverage of $\sim$ 620 \kms. The
frequency resolution was 2 MHz, which gives a velocity resolution of
5.2 \kms\ at 115 GHz; however, we have binned the data to a resolution
of 10.4 \kms\ to increase the signal to noise ratio.

For map making we used the data reduction software MIRIAD
\citep{sau95}. Data with unusually high visibilities were clipped, the
maps were naturally weighted and the inner quarter CLEANed to
1$\sigma$. The resulting synthesized beam is 3\farcs0$\times$2\farcs5
(PA=21$^\circ$) for NGC 2273 and 4\farcs5$\times$3\farcs0
(PA=$-$12$^\circ$) for NGC 5728. The rms noise in the maps is 3.09
Jy/beam \kms\ for NGC 2273 and 2.65 Jy/beam \kms\ for NGC 5728. The
maps cover a 60\arcsec$\times$60\arcsec\ area, but only the regions
with significant emission are shown in Figures \ref{intensity} and
\ref{intensity5728}.

\section{Molecular Gas Distribution, Dynamics, and Comparisons to Models \label{morphanddyn}}

\subsection{Morphology \label{morph}}

Figure \ref{intensity} shows the NIR image (top panel) of \citet{mul97}
as well as the integrated intensity \twcooz\ map (bottom panel) for NGC
2273. The CO emission is integrated over the velocity range where we
see emission, namely from 1669 \kms\ to 1981 \kms. The
\twcooz\ integrated intensity map (Fig.~\ref{intensity}) shows a small
bar-like structure (P.A.~$\sim$40\arcdeg) that is approximately
perpendicular to the primary bar of the galaxy (P.A.~$\sim$115\arcdeg).
Comparing the size of the CO bar with the synthesized beam, we see that
the bar is resolved along its major axis, but may not be fully resolved
along its minor axis. The integrated intensity map also shows
finger-like structures protruding to the north and south of the CO
bar. Assuming trailing spiral arms, these fingers line up with the
leading edge of the primary galactic bar.

The integrated intensity map only shows a two-dimensional view. If we
want to see the details of how the molecular gas is distributed, we
need to use the velocity information. The channel maps
(Fig.~\ref{n2273chanmap}) show that the CO bar is actually three
dynamically separate clumps that merge into a bar-like structure when
we average the emission over all channels. In order to emphasize the
three individual clumps, we have used the {\tt clumpfind} algorithm
\citep{wil94} and plotted the individual clumps in Fig.~\ref{clumps}.
The two clumps near the ends of the CO bar are brighter than the
central peak, which suggests that molecular gas may be flowing in along
the primary bar, but is actually being collected into clumps which are
presumably the location of the ILR. Similar CO morphologies have been
seen in other barred galaxies such as the `twin peaks' galaxies of
\citet{ken92} and the nearby starburst galaxy M82
\citep[\eg,][]{she95}. We will discuss the importance of these
similarities in \S\ref{other}.

The NIR image of NGC 2273 (top panel Figure \ref{intensity};
\citet{mul97}) shows isophote twists in the inner 10$\times$10$''$
misaligned from the primary bar by $\sim$90\arcdeg. The \twcooz\
integrated intensity map (bottom panel; Figure \ref{intensity})
shows a nuclear bar-like structure that aligns with the position angle
of the twisted NIR isophotes. Since we observe a similar morphology in
the CO map and in the NIR image, it strongly suggests that the NIR
isophote twists are the result of a disk phenomenon (as predicted by
Models 1 and 2) rather than a bulge phenomenon (Model 3). The models
of \citet{sha93} predict that we should see the nuclear molecular
component leading the nuclear isophote twists observed in the NIR by up
to 20\arcdeg. The models of \citet{fri93} predict that the gaseous
nuclear bar leads the gaseous stellar bar by up to $\sim$20\arcdeg. In
NGC 2273, there is no clear evidence that the gaseous secondary bar is
leading the stellar secondary bar by any significant amount. However,
this observation does not rule out either model, since \citet{sha93}
and \citet{fri93} predict deviation angles between the gaseous and
stellar bars as small as 5\arcdeg, which is smaller than we can measure
accurately from the maps.

Figure \ref{intensity5728} shows the $I$-band image (top panel) of
\citet{pra99} as well as the integrated intensity \twcooz\ map (bottom
panel) of NGC 5728. The CO emission is integrated over the velocity
range of 2650 \kms\ to 2980 \kms. The \twcooz\ map of NGC 5728
(Fig.~\ref{intensity5728}) does {\it not} show an obvious nuclear bar
as seen in NGC 2273. It contains several individual clumps of emission
that seem to form an arc with a radius of 6\arcsec. If this arc is part
of a molecular ring, then the CO ring is not aligned with the galactic
center, nor is it aligned with the ring structure surrounding
ionization cones seen in the HST images by \citet{wil93}. The brightest
peak in the CO map of Fig. \ref{intensity5728} is located
$\sim$2\arcsec\ to the SW of the brightest peak seen in the VLA and
H$\alpha$ maps of \citet{sch88}. The smaller CO clumps that run
counter-clockwise to the SE of the brightest peak line up with the dust
lane that runs out of the nucleus and along the primary bar of the galaxy
\citep{sch88}. We reserve discussion of the physical interpretation of
these features for \S\ref{other5728}.

At first glance, it appears that our CO map is offset from the galaxy
center (marked with a `$+$' in Fig.~\ref{intensity5728}) determined by
\citet{sch88}. It should be noted that the spectral lines in the
nucleus of NGC 5728 cover such a large range of velocities that we have
missed velocities lower than 2630 \kms. The H$\alpha$ maps of
\citet{sch88} show that there are some velocities as low as 2600
\kms\ and as high as 3010 \kms\ in the nucleus of NGC 5728. The lowest
velocity regions are located to the north-east of the dynamical center,
so that if there is strong CO emission at 2600 to 2630 \kms, we will have
missed it in our maps, perhaps creating the observed non-symmetric
appearance. There is definitely evidence of bright CO emission near
the low velocity end of our spectrometer. It can be seen as a faint
peak in Fig.~\ref{intensity5728} at RA = 14\h42\m24\fs2, $\delta$ =
-17\arcdeg15\arcmin07\farcs6 (it is the NE clump that is used in the
dynamical mass calculation of \S\ref{dyn}).

To test whether the asymmetry seen in the CO map of NGC 5728 is caused
by the exclusion of the lowest 30 \kms\ of the spectral data, we
cropped the highest 30 \kms\ from the spectra and recreated the moment
maps. This process reduces the intensity of the brightest peak to the
west of the dynamical center, but the peak is still more than twice as
bright as the emission peak on the north-east side of the dynamical
center. Thus, it is likely that the asymmetric appearance of the CO
map is real, and not an artifact of missing low velocity emission.

The NIR image of NGC 5728 \citep[Figure \ref{intensity5728}; see
also][]{sha93} shows isophote twists in the nuclear region similar to
those of NGC 2273, but our CO map of NGC 5728 shows no clear evidence
for the existence of a nuclear bar. This result suggests that in NGC
5728, the NIR isophote twists may not be the result of the nuclear bar,
but may be caused by a triaxial stellar bulge as predicted by
\citet{kor79}. The asymmetric CO distribution may also be evidence for
a past merger event in NGC 5728. We give a more detailed discussion of the possible
causes of the NIR isophote twists and discuss causes for the asymmetric
gas distribution in \S\ref{other5728}.

\subsection{Dynamics \label{dyn}}

The position-velocity diagram (Fig.~\ref{posvel}) taken along the axis
of the CO bar in NGC 2273 shown in Fig.~\ref{intensity} indicates that
the bar has a velocity gradient of $\approx$ 600 \kms\ kpc$^{-1}$
($\sim 900$ \kms\ kpc$^{-1}$ deprojected). Since we have no detections
beyond the secondary bar shown in Fig.~\ref{intensity}, and there are no
rotation curves published for the inner 1$'$, we cannot determine yet
if the CO secondary bar is kinematically distinct as predicted by Model (2)
or if it is rotating at the same angular frequency as the primary bar as
predicted by Model (1). In addition, without published rotation curves
for the region just beyond the nuclear bar, we are unable to estimated
the nuclear bar pattern speed.

Assuming Keplerian rotation, we have $$M_{\rm dyn} = \left({V_{\rm
circ}\over{{\rm sin}~i}}\right)^2 \left({R\over{G}}\right)$$ where
$M_{\rm dyn}$ is the mass interior to radius $R$, $V_{\rm circ}$ is the
projected circular velocity of the material at radius $R$, $i$ is the galaxy's
inclination and $G$ is the gravitational constant. Contributions from
non-circular velocities can only cause deviations of approximately 30\%
in the calculated mass \citep[\eg,][]{sak99}. Using Figure \ref{clumps}
to obtain velocities at the bar ends, and assuming the inclination to
be 41$^\circ$, we find the dynamical mass of the nuclear molecular bar
to be 1.6 $\times$ 10$^9$ \msun.

The channel maps of NGC 5728 (Figure \ref{n5728chanmap}) show a much
less ordered appearance than in NGC 2273. Careful examination of the
velocity of the clumps suggests that the southern-most clump has the
highest recession velocity and the velocities of the clumps decrease as
you move clockwise around the arc (ignoring the CO emission associated
with the dust lane to the far south). This is consistent with the
velocity field determined by the H$\alpha$ maps of \citet{sch88}. The
CO emission is very clumpy with some of the features seen in the
integrated intensity map of Figure \ref{intensity5728} being made up of
numerous CO features that are separated in velocity space. The
brightest peak is actually a superposition of many fainter peaks at
different velocities. There is no evidence in our CO maps of NGC 5728
for a bar interior to the ring as reported by \citet{pra99}, so
naturally we cannot confirm the reports that this secondary bar may be
counter-rotating.

Our CO data for NGC 5728 are not sufficient to create a high quality
position-velocity map as we did for NGC 2273, but we can use the
channel maps to determine the position and velocity for the two bright
clumps that lie nearly along the kinematic major axis of NGC 5728 and
straddle the dynamical center\footnote{The north east peak is at RA =
14\h42\m24\fs2, $\delta$ =-17\arcdeg15\arcmin07\farcs6 and has a
central velocity of 2660 \kms. The south west peak is at RA =
14\h42\m23\fs75, $\delta$ = -17\arcdeg15\arcmin13\farcs0 and has a
central velocity of 2960 \kms. These points are indicated in Figure
\ref{intensity5728} by $\times$'s.}. Using the same equations as for
NGC 2273, we calculate a dynamical mass of 6.3$ \times 10^9$ \msun\ for
the inner 8\arcsec\ of NGC 5728. We note that since there is no strong
evidence for circular motion in the CO maps, there is extremely high
uncertainty associated with this value. An alternative value for the
dynamical mass of the inner kiloparsec of NGC 5728 is given in the next
section.

\section{Molecular Gas Mass and Gas Mass Fraction \label{mass}}

We use the CO flux over the entire nuclear bar of NGC 2273 to estimate
the molecular gas mass in the nuclear region. We adopt $$M_{\rm gas} =
1.6 \times 10^4 \left({D\over{\rm Mpc}}\right)^2 \left({S_{\rm
CO(1-0)}\over{\rm Jy~km~s^{-1}}}\right) \left({X\over{X_{\rm
Gal}}}\right)$$ \citep{wil95}, where $D$ is the distance to the galaxy,
$S_{\rm CO(1-0)}$ is the \twcooz\ flux, $X$ is the CO-to-H$_2$
conversion factor compared to the Galactic value ($X_{\rm Gal}$). The
constant at the start of the equation contains a factor of 1.36 to
account for other elements besides hydrogen. For lack of evidence to
the contrary, we adopt the Galactic value of the CO-to-H$_2$ conversion
factor of $3\times10^{20}$ cm$^{-2}$ (K \kms)$^{-1}$ \citep{sco87} for
the galaxies studied in this paper. The total CO flux for the nuclear
CO bar of NGC 2273 is 48.6 Jy \kms, which corresponds to a molecular
mass of 4.7$ \times 10^8$ \msun. If we repeat this calculation using
only the region interior to the locations used in the dynamical mass
calculation ($\sim$ 1 kpc along the length of the CO bar) we measure a
flux of 38.0 Jy \kms\ (uncorrected for primary beam fall off), which
corresponds to a molecular mass of 3.6$ \times 10^8$ \msun. This mass
constitutes approximately 20\% of the galaxy's dynamical mass within
this radius (\S\ref{dyn}).
 
Of course, the estimate for the molecular gas mass is only a lower
limit since the interferometer is insensitive to the large scale
structure that may be present in the nuclear regions of this barred
galaxy. Single dish \twcooz\ spectra of the inner 55\arcsec\ of NGC
2273 taken by \citet{you95} show a flux of 137 $\pm$ 24 Jy \kms,
suggesting that our interferometry maps may be missing up to 65\% of
the CO emission. Also important is the effect of the uncertainties in
the individual measurements used to calculate this ratio. For example,
in \S\ref{dyn}, if we assume that $V_{\rm circ}$, $i$, and $R$ each
have a conservative $\sim$10\% uncertainty, this results in a
$\sim$25\% uncertainty in $M_{\rm dyn}$. Similarly for the molecular
gas mass, if we assume a $\sim$10\% uncertainty for each measured
value, we obtain an uncertainty in $M_{\rm gas}$ of $\sim$20\%. The
final value for the ratio $M_{\rm gas}/M_{\rm dyn}$ is (20$\pm$6)\%.

\citet{yan99} find that the nuclear ring of NGC 2273 is very dusty and
calculate the mass of this dust to be $\sim 10^5$ \msun. Comparing this
to our molecular gas mass, we find that the gas-to-dust ratio is
approximately 3600 in the center of this Seyfert 2 galaxy. This ratio
is higher than found for nearby spiral galaxies \citep[M$_{\rm
gas}$/M$_{\rm dust} \sim$1000;][]{dev90}. It may not be surprising that
the ratio of gas to dust is higher in the center of this barred galaxy,
where we might expect an increase of nuclear molecular gas (caused by
inflow along the bar) as well as a decrease in dust due to the strong
ionizing nuclear source \citep{yan99}.

The CO flux for the inner 8$\times$8\arcsec\ of NGC 5728 is 25.6 Jy
\kms\ (uncorrected for primary beam fall off), which indicates a
molecular mass of $5.7 \times 10^8$ \msun. Using our dynamical mass
data we find that the molecular gas in the nucleus of NGC 5728
constitutes 9\% of the total mass. We believe that our determination of
the dynamical mass in such a complicated environment is likely
unreliable. Our dynamical mass determined in \S\ref{dyn} is lower than
the value determined by \citet{rub80} through model fits of the
velocity data over the entire galaxy. Due to the complexity of our data
and the simplicity of the models assumed in \S\ref{dyn}, we will adopt
the value determined by \citet{rub80}. She finds the total dynamical
mass of the galaxy in the inner 10$''$ diameter (1.8 kpc) is $\sim
1\times 10^{10}$ \msun. This estimate lowers the mass fraction of
molecular gas to (6$\pm$2)\%. Again, this result is a lower limit because
the interferometer will miss large scale structure. There are no single
dish \twcooz\ spectra for NGC 5728 published, so we cannot directly
estimate the flux missed by our interferometric maps. Single dish
\twcoto\ spectra of the inner 22\arcsec\ of NGC 5728 (Petitpas \&
Wilson, in prep.) show fluxes of 62 Jy \kms. Assuming a
\twcoto/\joz\ flux ratio of 2.8 \citep{sak95} we obtain a \twcooz\ flux
of 22 Jy \kms. Comparing this to the CO flux in the interferometric map
of 30 Jy \kms\ for this region, we find that the interferometric maps
are detecting all of the flux detected by a single dish.

Both the models of \citet{sha93} and \citet{fri93} need to contain
molecular gas to create and sustain nuclear bars and twists.
\citet{fri93} found that they could create nuclear bars if the
{\it total} (atomic + molecular) gas to total mass ratio is 10\%, or if
2\% of the total mass interior to 1 kpc (diameter) is molecular.
\citet{sha93} were less specific and only state that 4-6\% of the
entire galaxy mass needs to be gaseous; a large portion of this gas is presumably
transferred inward at later stages, but specific values are not
quoted. The high gas contents in these two models are required to
provide enough dissipation so that the nuclear ring can become phase
shifted out of its stable orbit and collapse into a nuclear bar or
twist out of alignment. Both galaxies meet the initial nuclear
molecular gas mass requirements of \citet{fri96} yet only NGC 2273
shows evidence for a nuclear molecular bar. It is interesting that we
find $\sim$20\% gas (by mass) in the nucleus of NGC 2273, where we see
a nuclear bar, yet we only see $\sim$6\% (by mass) in the nucleus of
NGC 5728, where we see no evidence in our CO maps for a nuclear bar. We
defer discussion on potential reasons for the differences for
\S\ref{prev} and \ref{other}.

A recent interferometric \twcooz\ survey by \citet{sak99} suggests that
typical gas mass fractions range from 0.9\% to 31.5\% in the inner 500
pc radius of a sample of 17 galaxies, with no apparent correlation with
galaxy type. Since we have used the same techniques to determine the
gas mass fraction, it is worthwhile to compare our results with those
of a larger sample. For NGC 2273, we calculate the molecular gas mass
fraction over the entire length of the bar for which we see emission in
the position velocity plot (Figure \ref{posvel}), which corresponds to
the inner 720$\times$720 pc. For NGC 5728, we measured the molecular
gas mass fraction over an area of 8$\times$8\arcsec\ which corresponds
to the inner 1.4$\times$1.4 kpc. NGC 2273 has a nuclear gas mass
fraction of 20\% and NGC 5728 has 6\% molecular gas. \citet{sak99} find
evidence that HII nuclei have higher ratios of gas to dynamical masses
($\sim$18\% average) compared to Seyfert and LINER galaxies ($\sim$6\%
average). NGC 5728 is in agreement with the previously observed range
for its type of nuclear activity, while NGC 2273 seems to have a higher
molecular gas fraction than the other Seyfert 2 galaxies in the sample
obtained by \citet{sak99}. A distinct classification of NGC 2273 into
one particular category is difficult in light of observations
which suggest that, in addition to its Seyfert 2 activity, NGC 2273 is
undergoing nuclear star formation \citep{mul96}. These observations
suggest that NGC 2273 may be an intermediate object that should fall
somewhere between the HII and Seyfert 2 classifications of
\citet{sak99}, in which case it is in the expected intermediate range
of gas mass fraction for such an object. In any case, our observed gas
mass fraction for NGC 2273 does not compromise the results of
\citet{sak99}.

Thus far we have concentrated our comparisons predominantly on the
models of double barred galaxies that are comprised of both stars and
gas. There is another class of models that attempt to explain nuclear
NIR isophote twists using purely stellar orbits \citep{mac00}. It is
known that there are different classes of orbits in a barred
potential.  The two important ones are the $x_1$ family that runs
parallel to the bar major axis, and the $x_2$ family, that runs
perpendicular to it \citep[\eg,][]{ath92}. It was thought that the
$x_2$ orbits of the large scale bar near the nucleus could form the
$x_1$ orbits of the smaller nuclear bar and the corotation radius of
the nuclear bar could correspond to the ILR of the large scale bar. In
this picture, the nuclear bar must {\it always} be aligned
perpendicular to the primary bar. \citet{fri93} rule out this model by
studying a large sample of double-barred galaxies, since they found
that not all the observed offset angles between the nuclear bar and the
primary bar can be explained by inclination effects. Recently,
\citet{mac00} find that there exist orbits in which particles in a
double-barred potential remain on closed orbits and may form the
building blocks of long-lived double-barred galaxies without the need
for a gaseous component. The particle nature of this model renders it
untestable with our molecular gas observations, since molecular gas
will not orbit in a particle-like manner. We mention it
here solely as a possible explanation as to why NGC 5728 shows a
nuclear stellar bar in the NIR images, but no nuclear bar in the CO
maps.

\section{Inflow and Gas Consumption in NGC 2273 \label{inflow}}

Since nuclear bars are believed to be responsible for the inflow of
molecular gas, we may expect to see signs of inflow in the CO
observations of NGC 2273. In Figure \ref{n2273moment1} we show the
velocity field (moment 1 map) for NGC 2273 superimposed on the
integrated intensity CO map. The lines of constant velocity start at
1700 \kms\ (the broken contour in the south-east) and continue up to
1980 \kms\ (the very short, western-most contour) in increments of 20
\kms. The white line is the 1 $\sigma$ integrated intensity contour
from Figure \ref{intensity}.

The velocity field does not show evidence of the characteristic
`S'-shaped contours indicative of inflowing molecular gas. One possible
explanation for this is that we are probably not resolving the bar
across the minor axis. The contours are nearly parallel across the
major axis and evenly spaced, which is consistent with solid body
rotation. The kinematic major axis of the nuclear bar aligns with the
kinematic major axis of the rest of the galaxy as seen in the HI maps
of \citet{van91}. A signature similar to this is seen in the
unresolved nuclear gas bar of NGC 3351 \citep{dev92}. Another
explanation for the lack of kinematic signs of inflow is that the
molecular gas may be contained solely within the three independent,
gravitationally bound clumps seen in Figure \ref{clumps}. If this is
the case, then there would not be sufficient inflowing material to be
detected by our interferometric observations.

If the molecular material is confined to individual clumps surrounding
the nucleus, we might ask if the dynamical drag on these clumps may be
an effective means of transporting molecular material deep into the
nucleus. The masses of the individual clumps are on the order of $6
\times 10^{7}$ \msun, much smaller than the dynamical mass of
the galaxy interior to these clumps ($\sim 1.5 \times 10^{9}$ \msun;
from Figure \ref{clumps}), which suggests that dynamical friction
resulting from the background stellar population may be an effective
means of transporting material into the nuclear regions. The dynamical
friction timescale may be written $$ f_{\rm fric} = {{2.64 \times
10^{8}}\over{ln~\Lambda}} \left ({r\over{\rm pc}}\right )^2 \left
({v\over{\rm km~s^{-1}}} \right )\left ({{\rm M_\odot}\over{M}} \right
)$$ where we have assumed an isothermal sphere density distribution and
$r$ and $v$ are the orbital properties of a cloud with mass $M$.
$\Lambda$ is the Coulomb logarithm and can be expressed $$ \Lambda =
230 \left ({b\over{\rm pc}} \right ) \left ({v\over{\rm
km~s^{-1}}}\right )^2 \left ({{\rm M_\odot}\over{M}} \right ) $$
\citep{bin87,jog99} where $b$ is the effective radius of the cloud from
the galaxy center.

For the two outer clumps in the nuclear gaseous bar of NGC 2273 (masses
$\sim 6 \times 10^{7}$ \msun) moving with an average velocity of 180
\kms\ (deprojected) at a radius of 200 pc (the average of the
outer-most clumps in Figure \ref{clumps}) we estimate the timescale for
dynamical friction to be $10^{7}$ years. This timescale is shorter than
the predicted lifetimes of nuclear bars \citep[$\sim$ few $\times
10^{8}$ yrs;][]{fri93} which suggests that dynamical friction may be an
effective means of transporting molecular gas into the nucleus of NGC
2273 and will likely in turn affect the evolution of the nuclear bar in
this galaxy. The combined mass of the two outer CO clumps is
approximately $1 \times 10^8$ \msun. Assuming that all of this gas is
transported inwards though dynamical friction, we estimate the inflow
rate to be on the order of 10 \msun/yr. We note that this value will be
modified if in fact we are not resolving the CO emission. Smaller
clumps will have smaller masses and thus longer timescales. Conversely,
clumps at smaller radii will have much shorter timescales, so if in
fact the individual clumps used in this analysis are actually comprised
of smaller clumps at a range of radii, our estimate of the dynamical
timescale will have to be altered accordingly.

It is interesting to compare the above inflow estimate with the star
formation rates predicted from the infrared luminousity of the nuclear
starburst in NGC 2273. Following \citet{hun86} and assuming the
infrared luminousity \citep[log $L_{FIR}/L_\odot$ = 10.11;][]{bon97}
comes from dust heated by the nuclear star formation seen in the
H$\alpha$ maps of \citet{san00}, we estimate star formation rates in
the nuclear bar of NGC 2273 to be $\sim$ 3 - 6 \msun\ yr$^{-1}$. This
value is in good agreement with the inflow rates derived from dynamical
friction arguments. Assuming that all the molecular gas detected with
the interferometer (namely, the gas in the dense clumps) is used, there is
enough material to feed these star formation rates for 10$^8$ years.
This lifetime is approximately the same or shorter than the predicted
lifetime of the nuclear bars in the simulations of \citet{fri93},
indicating that star formation will play an important role in the
evolution of double barred galaxies.

The star formation and gas inflow rates in NGC 2273 are very similar to
the values obtained for NGC 2782 using similar techniques
\citep{jog99}. Both the high star formation rates and high inflow rates
in these two galaxies which contain both nuclear gaseous and nuclear
stellar bars suggest that gas rich nuclear bars are a very effective
means of transporting gas into the nuclear regions where it may fuel
star formation and perhaps other forms of nuclear activity. We note as
a counter example the double barred galaxy NGC 4736 \citep{won00}. It
contains both a nuclear stellar and gaseous bar (although, somewhat
weaker than NGC 2273), yet it has a nuclear star formation rate of only
0.2 \msun\ yr$^{-1}$. We will need more observations of similar
galaxies to build up a large enough sample to draw any strong general
conclusions.

\section{Comparison to Observations at Other Wavelengths \label{prev}}

\subsection{NGC 2273 \label{other2273}}

It has long been known that large scale bars are a good way to
transport material into the nuclear region of a galaxy
\citep[\eg,][]{com94} by allowing material to flow inward along the
leading edge of the bar. Indications of such an inflow along the large
scale bar are visible in the CO map of NGC 2273, seen as fingers
extending NW and SE from the top and bottom (respectively) of the CO
bar. This observation suggests that inflow along the primary bar of NGC
2273 is depositing material onto the ends of the nuclear bar seen in
the CO maps. This material would then be free to flow along the nuclear
bar into the central regions, and could possibly supply the nucleus
with enough material to sustain the observed Seyfert and starburst
activity. It is possible that the large scale bar will re-supply the
nuclear bar with molecular material, allowing it to live longer than
predicted in \S\ref{inflow} when we assumed that the material in
the bar was all that was available to the nuclear starburst.

The channel maps of NGC 2273 show that the bar seen in the integrated
intensity maps is actually composed of three separate structures. These
structures are barely spatially resolved in our maps (if at all), but
they are clearly resolved in velocity (as seen in
Fig.~\ref{n2273chanmap}). Similar nuclear CO distributions have been
seen previously in single barred galaxies such as NGC
5383 and M82 \citep{she00,nei98}. In NGC 5383, \citet{she00} see bright
CO emission from ``twin peaks'' that coincide with the ILR radius, as
well as a central concentration near the nucleus. High resolution
observations suggest that the nuclear structure of NGC 5383 is actually
a nuclear spiral when viewed with the Hubble Space Telescope
\citep{she00}, which casts some doubt on the identification of nuclear
bars with lower resolution, ground-based images.

HST images \citep{mal98} and narrow-band color maps \citep{yan99} of
NGC 2273 show that the nucleus contains a ring elongated in the same
direction as our nuclear CO bar. The HST (WFPC2, F606W) images show a
central bright spot and indications of what may be flocculent spiral
arms interior to the ring. The bright blobs at the ends of the nuclear
CO bar correspond with the edges of the ring seen by \citet{mal98} and
\citet{yan99}. It is likely that this ring indicates the location of
the ILR of the primary bar, and thus, our nuclear bar ends occur at the
ILR radius as predicted by the models of \citet{fri93} and
\citet{sha93}.

\subsection{NGC 5728 \label{other5728}}

HST images of the nucleus of NGC 5728 show what appears to be an
off-center ring of young stars circling two ionization cones that seem
to originate from a point $\sim$3\arcsec\ west-north-west
($-$60\arcdeg) from the center of our maps \citep{wil93}. These images
also show a bar-like structure that is oriented at a position angle of
$\sim$90\arcdeg\ which \citet{wil93} suggest may be comprised of older
stars. Our CO map shows no bright emission peak at the vertex of the
ionization cones of the HST maps. The bright and faint CO peaks (used
in \S\ref{dyn} to determine the dynamical mass) that straddle the map
center are roughly perpendicular to the ionization cones. The other
peaks in the maps (curving to the south) are associated with the dust
lanes that run along the leading edge of the primary bar \citep{sch88}.
Our CO maps also do not show any strong evidence of a bar structure
oriented at PA = $\sim$90\arcdeg\ which might correspond to the bar
suggested by the NIR images of \citet{sha93}. If we assume that the
clumps (marked by $\times$'s in Figure \ref{intensity5728}) that
straddle the dynamical center (indicated by the $+$) correspond to
similar nuclear bar end enhancements as seen in NGC 2273 (Figure
\ref{intensity}) then the nuclear CO bar in NGC 5728 would be leading
the stellar nuclear bar seen in the NIR images by $\sim$45\arcdeg. This
angle is greater than the offsets predicted by \citet{sha93} and
\citet{fri93} for which a 20\arcdeg\ maximum offset is predicted. If
the nuclear stellar bar is counter rotating as reported by
\citet{pra99}, then the nuclear CO bar is trailing the primary bar, which
is only predicted in kinematically distinct nuclear bar models of
\citet{fri93}.

Unfortunately, the integrated intensity maps and the channel maps of
the nucleus of NGC 5728 show no {\it conclusive} evidence for a nuclear
bar in our data either aligned or misaligned with the NIR isophote
twists. This result suggests that if the NIR isophote twists of
\citet{sha93} are the result of a nuclear bar, it is mostly stellar and
contains little molecular gas. This suggests there is a possibility
that the molecular gas distribution has been modified by star formation
activity, and future models of double barred galaxies may need to
include the effects of star formation. This scenario is elaborated on
in the following section.

Another (more likely) possibility is that the off-center CO emission seen
in NGC 5728 may be evidence of a recent merger event. The lack of tidal
tails suggests that any merger experienced by NGC 5728 would have been
a minor merger. Simulations of minor mergers by \citet{her95} show
evidence for molecular gas off-set from the primary galaxy center. If
the asymmetric CO map is the result of a minor merger, it may be
responsible for triggering the large scale bar perturbation and could
account for the counter-rotating core reported by \citet{pra99}. Minor
mergers have been attributed as the possible cause of the `X-shaped' S0
galaxies \citep{mih95}, and the similarities between these X-structures
and the NIR isophote twists seen in NGC 5728 are noteworthy.

\section{Comparisons with Other Candidate Double Barred Galaxies \label{other}}


From the NIR surveys that have detected the isophote twists originally
thought to be the signature of a nuclear bar
\citep[\eg,][]{mul97,sha93,woz95,jar88,elm96,fri93,jog99} only a
handful of galaxies have high resolution CO maps published: NGC 2273,
NGC 5728, NGC 2782, NGC 4736, and NGC 6951 \citep[this
work;][]{jog99,sak99,ken92}. NGC 2273, NGC 2782, and NGC 6951 contain two
peaks at the ILR radius with more CO emission interior to these peaks.
CO maps of NGC 4736 show what may be a weak twin peak structure, with a
bright central concentration \citep{sak99}. NGC 5728 has a bright,
off-center peak of emission, with other emitting regions that do not
seem to line up with other features or the ILR radius. Only NGC 5728
does not show evidence for a twin peaked or bar structure elongated in
the same direction as the nuclear NIR isophote twists (possibly as
a result of the merger scenario discussed above).

To explain the variety of CO distributions seen in these galaxies,
we need to first consider the time evolution of the models. The models
of \citet{sha93} were only run until they reached a steady state, so
comparisons of our CO maps with the predictions for the lifetimes of
nuclear bars is not possible. \citet{fri93} show that after the nuclear
bar forms, it usually remains present for over 5 turns of the nuclear
bar (approximately 2 turns of the primary bar; $\sim$500 Myr). The
double-bar phase can transport the molecular gas inward, which
eventually results in the disruption of the nuclear bar; often even the
large-scale bar is disrupted by bulge growth \citep{pfe90} or central
mass concentration \citep{has90}. At all times in these simulations, the gas
distribution resembles the stellar distribution, even if they are
somewhat out of phase, so the models are incapable of explaining the observed
variety of CO distributions. Neither of the double barred galaxy models
by \citet{sha93} or \cite{fri93} include star formation, which is
abundant in this sample of galaxies and will likely have a strong
effect on the molecular gas properties within their nuclear regions.

If we consider the star formation activity of each of these galaxies,
we see that NGC 2273, NGC 2782, and NGC 6951 have active nuclear star
formation and the CO maps show a lumpy CO bar, a ordered CO bar, and a
distorted twin peak structure with faint central concentration,
respectively \citep[this paper;][]{jog99,koh99,ken92}. In NGC 4736, there no evidence
for active star formation in the nucleus, and the CO maps show a bright
central concentration \citep{kin93,sak99}. A summary of these
observations is given in Table \ref{gals}. These results suggest that
NGC 2273, NGC 2782, and NGC 6951 could perhaps be in later stages of
evolution, in which the nuclear molecular gas bar has been less
disrupted by star formation. NGC 4736 may not have yet undergone nuclear
star formation, so the molecular gas in the nucleus is not yet exhausted or disrupted. In addition, NGC 5728 shows evidence for past star
formation, but very little remains ongoing in the nucleus. If the
unusual gas distribution is not the result of a merger, but past rounds
of star formation, then NGC 5728 may have already disrupted or exhausted
the nuclear molecular gas and it has lost its nuclear gas barred
appearance, while the nuclear stellar bar remain unaffected. We
emphasize that the models of \citet{fri93} do not contain star formation
and thus show no evidence for such a time evolutionary sequence proposed here.
Paper II will explore the possible influence of star formation on the
molecular gas distributions in more detail. The accurate inclusion of star formation into the models will
help shed some light on the different CO morphologies observed in the
double barred galaxies.

Another explanation for the variety of CO morphologies may be that
there is different physics at work in the molecular gas in each
galaxy.  It is interesting that the NIR images of all five galaxies
look remarkably similar, while the CO maps look remarkably different.
The similarity in the NIR images suggests that the stellar distribution
in these galaxies is similar. The differences in the CO morphology
suggest that there may be something different about the molecular gas
in each galaxy, which causes it to respond differently to the galactic
stellar potential. We are in the process of analyzing multi-transition
CO data from in order to place constraints on the molecular gas
temperatures and densities for the galaxies discussed above. In Paper
II of this series, we will present evidence that the gas properties in
these galaxies may be responsible for the variety of CO distributions
seen.

In closing, we note that the identification of a gaseous nuclear bar
may be ambiguous: there may be confusion between what is considered a
``twin peak'' galaxy and what is considered a nuclear bar. Galaxies
with unresolved twin peaks will appear to have a nuclear CO bar. For
example, NGC 3351 is classified as a twin peak galaxy by \citet{ken92}
but considered to have a nuclear CO bar by \citet{dev92}. Twin peak
galaxies are believed to be caused by gas collecting at an ILR as it
flows inward along the leading edge of the primary bar of a galaxy
\citep{ken92}. These galaxies are thought to be {\it preventing}
molecular gas from reaching all the way into the nucleus, while nuclear
bar galaxies are thought to {\it assist} molecular gas flow into the
nucleus by transporting it interior to the ILR \citep{ken92,shl89}. We
will need a larger sample of galaxies containing unambiguous nuclear
bars so we can determine their properties in comparison to twin peak
galaxies.

\section{Summary \label{conc}}

We have mapped the barred galaxies NGC 2273 and NGC 5728 in
\twcooz\ with the Owens Valley Radio Observatory Millimeter Array. These galaxies are known to
contain NIR nuclear isophote twists, which are thought to be the
signature of a nuclear bar. The main results are summarized as
follows:

1) In NGC 2273 we see what appears to be a nuclear molecular bar
aligned with the nuclear stellar bar, both of which are approximately
perpendicular to the large scale galactic bar. Since they are both
perpendicular to the primary bar, we cannot rule out the possibility
that we are actually seeing gas and
stars populating the $x_2$-like orbits of the primary bar instead of a nuclear bar. The lack of a
detailed rotation curve for the inner regions of NGC 2273 prevents us
from determining if this nuclear bar is kinematically separate from the
rest of the galaxy.

2) Using dynamical friction arguments, we estimate the inflow rate
along the nuclear bar to be 6-10 \msun\ yr$^{-1}$. This high rate is
comparable to the value determined in the double barred galaxy NGC 2782
\citep{jog99} and suggests that dynamical friction may be an effective means of
transporting molecular gas in the nuclei of galaxies.

3) In the nucleus NGC 5728 we see a series of clumps of emission that
do not seem to align with any of the features previously observed at
other wavelengths. CO emission is detected coincident with the dust
lane to the south of the nucleus. The peak of the CO map is not aligned
with the galactic center, nor is it located at the center of the offset
nuclear ring seen in the HST images of \citet{wil93}. We do not see
evidence for a nuclear bar in the CO maps, which suggests that, if
there is a real nuclear bar in the NGC 5728, it must contain
little or no molecular gas. It is possible that the NIR isophote twists
thought to be the nuclear stellar bar in NGC 5728 are caused by either
a triaxial stellar bulge or are scattered light from the jets observed
in the HST images of \citet{wil93}. The asymmetric gas distribution
suggests that NGC 5728 may have undergone a minor merger event in its
past.

4) We have calculated the molecular gas mass over the entire nuclear
bar of NGC 2273 and find that it contains 4.7$\times10^8$ \msun\ of
molecular gas. The inner 1.4$\times$1.4 kpc of NGC 5728 contains
5.7$\times10^8$ \msun\ of molecular gas. Assuming Keplerian motion, we
find that the dynamical mass of the inner 10\arcsec\ of NGC 2273 is
1.9$\times10^9$ \msun, which translates into a molecular gas mass
fraction of 20\% for the nucleus of NGC 2273. Adopting the dynamical
mass value determined by \citet{rub80} we calculate the molecular mass
fraction to be 6\% in the nucleus of NGC 5728.

5) Both galaxies contain sufficient molecular gas as required by the
initial values of the models by \citet{sha93} and \citet{fri93} (this
is presumed to increase with time, but time dependent values are not
quoted). However, we only see a nuclear gas bar in NGC 2273, possibly
because of the potential merger history of NGC 5728.

6) The similarity in the NIR images and the differences in the CO maps
accumulated for a sample of galaxies with NIR isophote twists suggest
that the molecular gas may have different physical properties or is modified by
star formation, which allows it to respond differently in similar gravitational
potentials. Multi-wavelength
data have been obtained to test this hypothesis and will be
presented in Paper II.

\acknowledgments 

This research has been supported by a research grant to C.~D.~W.~from
NSERC (Canada). This research has made use of the NASA/IPAC
Extragalactic Database (NED) which is operated by the Jet Propulsion
Laboratory, California Institute of Technology, under contract with the
National Aeronautics and Space Administration. The Owens Valley Radio Observatory Millimeter Array is operated by the California Institute of Technology and is
supported by NSF grant AST 96-13717.

\clearpage

\figcaption[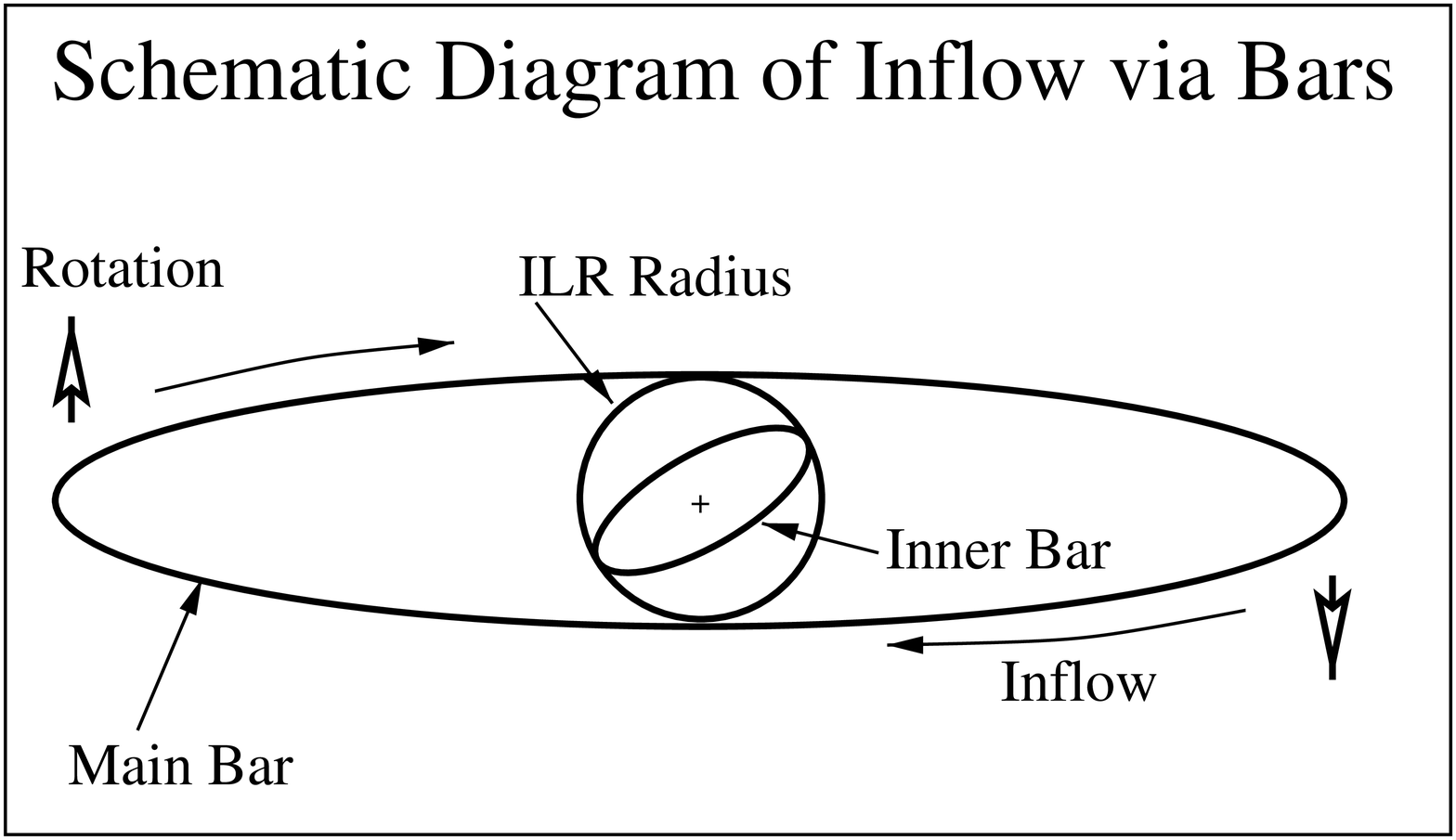]{Schematic diagram of how a large scale bar
(Main Bar) passing though a disk can transport material towards the
nucleus of the galaxy. It is difficult for simulations to transport
material all the way to the center because it tends to get trapped at
the Inner Lindblad Resonance (ILR). This figure shows the Inner Bar
that may form interior to the ILR and transport material all the way to
the nucleus. 
\label{schematic} 
}

\figcaption[f2.eps]{Integrated intensity map of \twcooz\ emission in
NGC 2273. The CO map of NGC 2273 (bottom panel) shows a nuclear bar
aligned with the NIR isophote twists of Mulchaey \etal\ (1997; top
panel). The fingers of emission to the north and south of the nuclear
CO bar suggest evidence of inflow onto the secondary bar along the leading
edge of the primary bar. In the NIR image the contours start at 17 mag
arcsec$^{-2}$ (K$_S$ band) and are in 0.5 mag increments. The rms noise
in the CO map is 3.1 Jy/beam \kms\ and the contour levels for the CO
integrated intensity map start at 1$\sigma$ and increase in steps of
1$\sigma$. The beam size of 3$'' \times 2.5''$ with a PA of
21\arcdeg\ is shown in the lower left corner.
\label{intensity} 
}

\figcaption[f3.eps]{Channel maps of CO emission in NGC 2273. The CO bar
seen in Fig.~\ref{intensity} is actually comprised of three clumps. The
contour levels are 0.04, 0.06, 0.08 ... Jy/beam (2, 3, 4 ... $\sigma$).
The plus sign indicates the map center. The panels are shown at 31.2
\kms\ intervals and are binned to 31.2 \kms\ wide bins.
\label{n2273chanmap}
}

\figcaption[f4.eps]{Clumps of CO emission that form the nuclear bar
of NGC 2273. This figure shows the output from the {\tt clumpfind}
algorithm. It shows the three main clumps of CO emission that combine
to produce the bar seen in the integrated intensity map
(Fig.\ref{intensity}). \label{clumps}
}

\figcaption[f5.eps]{\twcooz\ integrated intensity map for the nucleus
of NGC 5728 (bottom panel). Note that we do {\it not} see an secondary gas
bar as we do for NGC 2273. The top panel shows the composite I-band
image from \citet{pra99} indicating the location of the nuclear stellar
bar. The CO emission appears be asymmetric with respect to the
dynamical center of the galaxy (marked by the $+$) determined by
\citet{sch88}. The rms noise in the map is 2.65 Jy/beam \kms, and the
contours start at 0.5$\sigma$ and increase in steps of 0.5$\sigma$. The
beam size is $4.5'' \times 3''$ with a PA=$-$12\arcdeg\ and is shown in
the lower left corner. The $\times$ symbols indicate the clumps used to
determine the dynamical mass in \S\ref{dyn}.
\label{intensity5728}
}

\figcaption[f6.eps]{Position-velocity diagram for a slice along the
major axis of the CO bar in NGC 2273 shown in Fig.~\ref{intensity}. It
rotates as a solid body with a projected velocity gradient of 570
km/s/kpc. Contours indicate 10\%, 20\%, 30\% ... of the peak. Positive
offsets are toward the north-east end of the bar.
\label{posvel}
}

\figcaption[f7.eps]{Channel maps of NGC 5728. Note how there is
not strong evidence for large scale ordered motion. The large plus sign
indicates the galaxy center determined by \citet{sch88}. The smaller
crosses mark the locations of the peaks used to calculate the
dynamical mass. Contours levels are 0.04, 0.06, 0.08, ... Jy/beam (1.5,
2.3, 3 ... $\sigma$). The panels are shown at 20.8 \kms\ intervals and
are binned to 20.8 \kms\ wide bins.
\label{n5728chanmap}
}

\figcaption[f8.eps]{Moment 1 map for the CO bar of NGC 2273. The black
lines are lines of constant velocity, ranging from 1700 \kms\ to 1960
\kms\ in increments of 20 \kms. The underlying grayscale is the CO
integrated intensity map and the white contour is the 1 $\sigma$
contour from Figure \ref{intensity}. 
\label{n2273moment1}
}

\clearpage

\begin{deluxetable}{lccl}
\tablecaption{Adopted Properties of NGC 2273 and NGC 5728 \label{table}}
\tablehead{
\colhead{Property} & \colhead{NGC 2273} & \colhead{NGC 5728} & \colhead{References} \\
          } 
\startdata
R.A. (2000.0)         & 06\h50\m08\fs7 & 14\h42\m24\fs0  & 1 \\
Dec. (2000.0)         & +60\arcdeg50\arcmin45\farcs1 & $-$17\arcdeg15\arcmin10\farcs8   & 1 \\
Classification        & SB(r)a & SAB(r)a  & 2 \\
Optical Diameter      & 3\farcm3 & 3\farcm1 & 1, 2 \\
Nuclear Ring Diameter & $\sim$10\arcsec & $\sim$10\arcsec & 3, 4 \\
Outer Ring Diameter(s)& 0\farcm7, 2\farcm2, 3\farcm1 & 55\arcsec & 4, 5 \\
Inclination           & 41\arcdeg & 48\arcdeg & 2, 4 \\
Primary Bar PA           & $\sim$115\arcdeg & $\sim$38\arcdeg & 4, 6 \\
NIR Isophote Twist PA & $\sim$45\arcdeg & $\sim$90\arcdeg & 6, 7 \\
Heliocentric Velocity & 1841\kms & 2788\kms & 2 \\
Assumed Distance      & 24.5 Mpc & 37.2 Mpc & 8 \\
Linear Scale          & 1\arcsec\ = 120 pc & 1\arcsec\ = 180 pc & 8 \\
\enddata 
\tablerefs{(1) NASA/IPAC Extragalactic Database; (2) \citet{dev91}; (3)
\citet{yan99}; (4) \citet{sch88}; (5) \citet{van91}; (6) \citet{mul97};
(7) \citet{sha93}; (8) Assumes H$_0$ = 75 \kms\ Mpc$^{-1}$ }

\end{deluxetable}

\clearpage
\begin{deluxetable}{lcp{4cm}p{4cm}l}
\tablecaption{Properties NIR Isophote Twist Galaxies \label{gals}}
\tablehead{
\colhead{Galaxy} & \colhead{M$_{gas}$/M$_{dyn}$} & \colhead{Nuclear} & \colhead{Nuclear CO} & \colhead{References} \\
\colhead{~} & \colhead{(nuclear)} & \colhead{Activity} & \colhead{Morphology} & \colhead{~} \\
          } 
\startdata
NGC 2273  & 20\%  &  \raggedright Star formation, Seyfert 2 activity & \raggedright Bar/triple peaks & 1, 2  \\
NGC 2782  &  8\%  &  \raggedright Star formation, Seyfert 1 activity & \raggedright Bar/twin peaks & 3  \\
NGC 4736  &  2\%  &  \raggedright  Star formation in ring outside nucleus, LINER  &   \raggedright Central peak, weak signs of twin peaks? & 4, 5  \\
NGC 5728  &  6\%  &  \raggedright  Evidence of past star formation, Seyfert 2  & \raggedright Non-centralized peaks, no bar & 1, 6  \\
NGC 6951  & 29\%  &  \raggedright  Star formation  &   \raggedright Twin (triple?) peak/ elongated ring? & 7  \\
\enddata 
\tablerefs{(1) this paper; (2) \citet{mul96}; (3) \citet{jog99} (4) \citet{sak99}; (5) \citet{kin93}; (6) \citet{wil93}; (7) \citet{koh99}  }

\end{deluxetable}

\clearpage

\plotone{f1.eps}
\plotone{f2.eps}
\plotone{f3.eps}
\plotone{f4.eps}
\plotone{f5.eps}
\plotone{f6.eps}
\plotone{f7.eps}
\plotone{f8.eps}

\end{document}